\documentclass[11pt]{article}
\usepackage{a4,amssymb,verbatim,psfig,amsmath}

\newtheorem{theorem}{Theorem}[section]
\newtheorem{lemma}[theorem]{Lemma}

\newtheorem{ex}{Example}[section]
\newenvironment{example}{\begin{ex}\rm}{ \hfill $\Diamond$ \end{ex}
        \vskip4pt}             
\newtheorem{ass}{Assumption}[section]
             
\numberwithin{equation}{section}

\begin{document}
\def\dx#1{{\partial \over \partial#1}}

\def\dh#1{\mathop {#1}\limits_{h}}

\def\dhdva#1{\mathop {#1}\limits_{{h \over 2}}}

\def\dhp#1{\mathop {#1}\limits_{+h}}
\def\dhpi#1{\mathop {#1}\limits_{+{h_i}}}
\def\dhm#1{ \mathop{#1}\limits_{-h}}
\def\dphh#1{ \mathop{#1}\limits_{h \bar h}}
\def\da#1{ \mathop{#1}\limits_{+\tau}}
\def\db#1{ \mathop{#1}\limits_{-\tau}}
\def\dc#1{ \mathop{#1}\limits_{\pm \tau}}
\def\dd#1{ \mathop{#1}\limits_{+h}}
\def\df#1{ \mathop{#1}\limits_{-h}}
\def\dpm#1{ \mathop{#1}\limits_{\pm h}}
\def\dg#1{ \mathop{#1}\limits_{\pm h}}
\newcommand{\ddt}{\partial \over \partial t}
\newcommand{\ddx}{\partial \over \partial x}
\newcommand{\ddy}{\partial \over \partial y}
\newcommand{\ddz}{\partial \over \partial z}
\newcommand{\ddyy}{\partial \over \partial y'}
\newcommand{\ddu}{\partial \over \partial u}
\newcommand{\ddv}{\partial \over \partial v}
\newcommand{\ddw}{\partial \over \partial w}
\newcommand{\ddui}{\partial \over \partial u^{i}}

\def\dhxpl#1{ \mathop{#1}\limits_{+h_{x}}}
\def\dhypl#1{ \mathop{#1}\limits_{+h_{y}}}
\def\dhzpl#1{ \mathop{#1}\limits_{+h_{z}}}
\def\dhxm#1{ \mathop{#1}\limits_{-h_{x}}}
\def\dhym#1{ \mathop{#1}\limits_{-h_{y}}}
\def\dhzm#1{ \mathop{#1}\limits_{-h_{z}}}

\def\dhx#1{ \mathop{#1}\limits_{h_{x}}}
\def\dhy#1{ \mathop{#1}\limits_{h_{y}}}
\def\dhz#1{ \mathop{#1}\limits_{h_{z}}}

\begin{center}
{\Large {\bf Conservation laws of  semidiscrete \\
  $  $ \\ 
canonical Hamiltonian equations} }
\end{center}

\bigskip

\begin{center}
{\large  Roman Kozlov } 

\bigskip

Department of Mathematical Sciences, NTNU, N-7491,
Trondheim, Norway 

\end{center}

\bigskip

\begin{abstract}
There are many evolution partial differential equations which can 
be cast into Hamiltonian form. 
Conservation laws of these equations are 
related to one--parameter Hamiltonian symmetries 
admitted by the PDEs~\cite{[2]}. The same result 
holds for semidiscrete Hamiltonian equations~\cite{[Kozl1]}. 
In this paper we consider semidiscrete canonical Hamiltonian equations. 
Using symmetries, we find conservation laws for the semidiscretized 
nonlinear wave equation and Schr\"odinger equation. 
\end{abstract}

\section{Introduction}

Many partial differential equations (PDEs) of nondissipative 
continuum mechanics can be presented in Hamiltonian form 
(see~\cite{[Marsd]} and references therein). 
It is well known that conservation laws of Hamiltonian PDEs 
are related to 
one--parameter Hamiltonian symmetries~\cite{[2]}. 
The analog of this result holds in the semidiscrete case 
with no change in the statement although the framework 
must be modified~\cite{[Kozl1]}.

In the present paper we examine canonical Hamiltonian equations
\begin{equation} \label{equ5} 
{ \bf v } _t = { \delta {\cal H } \over  \delta { \bf  w} } , \qquad 
{ \bf w } _t = - { \delta {\cal H } \over  \delta { \bf  v} }  , 
\end{equation}
which form a special type of the Hamiltonian equations. 
For instance, the nonlinear wave equation 
\begin{equation} \label{EQ1} 
v_{tt} = \Delta v  - V'(v), 
\end{equation}
and the nonlinear Schr\"odinger equation
\begin{equation} \label{EQ2} 
i \psi _t + \Delta \psi + F' ( |\psi|^2 ) \psi = 0 . 
\end{equation}
can be rewritten in the form (\ref{equ5}). 
We will introduce semidiscrete analogs of the 
canonical Hamiltonian PDEs and show how one can use 
the Hamiltonian form of Noether's theorem 
to find conservation laws of these equations.

To find conservation laws of semidiscrete equations with the help 
of Noether's theorem, we need to know symmetries 
in the evolutionary form. 
Different methods which can be used to find symmetries of discrete and 
semidiscrete equations are discussed, for example, in~\cite{[3]}. 
The most successful application 
was shown for linear difference 
equations~\cite{[3]},\cite{[Flore]},\cite{[Vinet]},  
where for the considered equations 
symmetry algebras isomorphic to those of the underlying 
continuous equations were found. 
However, it is not known how to find all symmetries admitted by 
nonlinear semidiscrete equations.

One of the possibilities, which we exploit in this paper, is 
to make use of the admitted Lie point symmetries. 
Many Lie point symmetries of semidiscrete equations 
can be easily found as 
Lie point symmetries of the underlying 
continuous equations 
preserved under the space discretization. 
In both continuous and semidiscrete cases 
these symmetries are given by the same vector 
fields. Using factorization, one can obtain 
corresponding evolutionary operators. 
If the latter are Hamiltonian symmetries, 
they let us find conservation laws. 
Obviously, for our purpose we will not be interested 
in all symmetries but only in Hamiltonian ones.

The layout of the paper is as follows: 
In Section~2 we briefly introduce Hamiltonian equations 
and specify canonical Hamiltonian equations. 
Symmetries and the Hamiltonian form of Noether's theorem are 
discussed in Section~3. In Sections~4 and~5 we examine 
semidicretizations of equations (\ref{EQ1}) and (\ref{EQ2}) 
and find their conservation laws. In final Section~6 we 
make concluding remarks. In particular, we mention the connection 
between Euler--Lagrange equations and canonical 
Hamiltonian equations.

\section{Hamiltonian equations} 

For simplicity we will consider the case of 
one space coordinate $x$. We assume that the solutions 
are sufficiently smooth, 
all variational derivatives tend to zero as 
the solution tends to zero and 
the solution  and a number of its space derivatives tend to zero 
as $|x| \rightarrow \infty $. 
We suppose that the solution decays fast enough so that 
all integrals and sums make sense. 

\subsection{Hamiltonian partial differential equations}

Many systems of evolution equations 
$$
{ \bf u}_t  = K ( x, {\bf u}^{(m)} )  , 
$$
where $ { \bf u}$ stands for $N$ dependent variables 
$ { \bf u } = ( u^1, u^2, ..., u^N)^T $ 
and  
${\bf u}^{(m)} = ({ \bf u}, {\bf u}_1, {\bf u} _2, ..., { \bf u} _{m} ) $ 
represents ${ \bf u} $ and a finite set of derivatives of ${ \bf u} $ 
with respect to space coordinate $x$, 
can be  cast into the Hamiltonian form
\begin{equation} \label{eqy2}
{ \bf u}_t = {  \cal D} 
\left( { \delta {\cal H}  \over \delta { \bf u} } \right) , \qquad 
{\cal{H}} [{\bf u}] = \int H(x, { \bf u} ^{(n)}) dx, \qquad 
{ \delta H \over \delta { \bf u} } = 
\left(  { \delta H \over \delta u^1 },  { \delta H \over \delta u^2 }, ..., 
 { \delta H \over \delta u^l } \right) ^T 
\end{equation}
with the help of the Hamiltonian functional $ {\cal{H}} [{\bf u}] $, 
variational operator  $ \delta \cdot  / \delta { \bf u} $  and 
the linear operator $ {\cal{D}} $~\cite{[2]}. 
We denote as ${\cal F}$ the space of functionals 
$$
\int P (t, x, {\bf u}^{(k)}) dx, \quad k \in  \mathbb{N}.   
$$
The operator $ {\cal{D}} $ must be Hamiltonian, i.e., it forms 
the Poisson bracket 
\begin{equation}  \label{brac1} 
\{  {\cal{P} }  , {\cal{L} } \} = 
\int  
\left(  { \delta {\cal{P}} \over \delta {\bf u}  } \right)^T 
{\cal{D}} 
\left(  { \delta {\cal{L}} \over \delta {\bf u}  } \right) dx 
\end{equation}
satisfying the conditions of {\it skew-symmetry} 
\begin{equation} \label{cond2} 
 \{  {\cal{P}}  , {\cal{L} } \} = 
- \{  {\cal{L}}  , {\cal{P} } \} 
\end{equation}
and the {\it Jacobi identity } 
\begin{equation} \label{cond3} 
\{ \{  {\cal{P}}  , {\cal{L} } \} ,  {\cal{R} }  \} + 
\{ \{  {\cal{R}}  , {\cal{P} } \} ,  {\cal{L} }  \} + 
\{ \{  {\cal{L}}  , {\cal{R} } \} ,  {\cal{P} }  \} = 0 
\end{equation}
for all functionals $\cal{P}, \cal{L}, \cal{R} \in {\cal F}$.  

The variational derivatives of a functional 
can be found by the action of the Euler operators  on the 
integrand
\begin{equation} \label{equ11} 
\begin{array}{c}
{ \displaystyle 
{ \delta {\cal{H}} \over \delta u^i   } = E^i  (H) , \qquad 
i = 1, ..., N , } \\
\\
{ \displaystyle 
E^i  = { \partial \ \cdot \over \partial u^i  } - 
 D_x \left( { \partial \ \cdot \over \partial u^i_1 } \right) + 
D_x^2 \left( { \partial \ \cdot \over \partial u_{2}^i } \right) + 
\cdots + 
(- D_x )^n \left( { \partial \ \cdot \over \partial u_{n}^i } \right)
+ \cdots   , } 
\end{array} 
\end{equation}
where $D_x$ is the total space derivative operator. 

\subsection{Canonical Hamiltonian equations}

Canonical Hamiltonian equations 
form a subset of the equations (\ref{eqy2})    
characterized by an 
even--dimensional space of dependent variables 
$N = 2n$, ${ \bf u} =  (  v^1 , ..., v^n , w^1 , ..., w^n )^T$, 
and the canonical Hamiltonian operator
\begin{equation} \label{eqtt5} 
J  = \left(  
\begin{array}{cc} 
0_n & I_n \\
-I_n  &  0_n  \\
\end{array} 
\right) , 
\end{equation}
where $I_n$ is the $n \times n$ identity matrix and 
$0_n$ is the $n \times n$ zero matrix. 
Thus these Hamiltonian equations have the form (\ref{equ5}). 
It is easy to see that the Poisson bracket generated by operator $J$ 
$$
\{  {\cal{P} }  , {\cal{L} } \} = 
\int  
\sum_{i=1} ^n 
\left( 
 { \delta {\cal{P}} \over \delta  v^i   }  
 { \delta {\cal{L}} \over \delta  w^i   } 
- 
 { \delta {\cal{P}} \over \delta  w^i   }  
 { \delta {\cal{L}} \over \delta  v^i   } 
\right) dx 
$$
satisfies skew--symmetry (\ref{cond2}) 
and Jacobi identity (\ref{cond3}).

\subsection{Semidiscrete Hamiltonian equations}

Given a Hamiltonian PDE, we attempt to discretize both 
the Poisson bracket and the Hamiltonian functional so that we 
preserve Hamiltonian structure. 

To consider semidiscrete equations we 
introduce a two--dimensional mesh which is uniform 
in space and continuous in time. 
Let us denote the mesh   points as
$\{ x_i(t) \}$, $i \in   \mathbb{Z}$, $t \geq 0 $ 
and define mesh  $\Omega$   by two conditions: 
\begin{equation}  \label{meshom}
\Omega: \qquad  
x_{i+1}( t)  - x_{i} ( t) =   x_{i}( t) - x_{i-1}( t)  ,\quad  
x_{i} ( t + \tau ) =  x_{i} ( t)  ,  \quad 
i \in   \mathbb{Z},  \quad t , \tau    \geq 0 . 
\end{equation} 
The first equation requires the space mesh to be uniform 
for any fixed time. 
The second equation requires that 
only vertical mesh lines in the time--space plane are considered.

Now we can introduce discrete space derivatives 
${  \dh u { } _1 ^{i}  }  = {\dd{D} } u ^ {i}$, 
${  \dh u { } _{2} ^ {i} }  = {\df{D}  } {  \dh   u { } _{1} ^ {i}  } $, ...,
${  \dh u { } _{2k+1 }^ {i}}  = {\dd{D}  }{   \dh  u { }_{2k}^ {i} } $, 
${  \dh u { }_{2k+2 }^ {i}}  
= {\df{D}  }  {  \dh u { }_{2k+1}^ {i}}  $, ..., $ i =1, ..., l$, where 
$ {\dd{D} } $ and $ {\df{D} } $ are the right and 
left discrete differentiation operators
\begin{equation}  \label{eqa2}
{\dd{D} }   = { S_{+}  - 1  \over h };\qquad 
 {\df{D} }   = { 1- S_{-}  \over h }, 
\end{equation}
defined with the help of 
the right shift $S_{+}$ and left shift $S_{-}$
operators 
\begin{equation} \label{eq24qq}
S_{+}  f(x) = f(x+h ) ,\qquad  S_{-}  f(x) = f(x-h ) . 
\end{equation}

We will consider the space of discrete derivatives 
${\dh { \bf u} } ^{(m)} 
= ( { \bf  u } , \dh { \bf u } { } _1  , \dh { \bf u} { } _2 , ..., \dh { \bf u} { } _m)$ 
and  functionals of the form 
\begin{equation} \label{equ14} 
\dh {\cal{P}}  = \sum_{\Omega}  
 { \dh P }  ( t, x , h, { \dh { \bf u } }  ^{(m)}  ) h ,
\end{equation}
where the summation is taken over all space points of mesh $\Omega$ 
for some  fixed time. 
We denote the space of such functionals as $ \dh {\cal F} $.

We assume that the Hamiltonian operator $ {\cal{D}}$ can be approximated by 
an operator $ \dh{ {\cal{D}} }$  such that the discrete bracket 
\begin{equation} \label{eq2}
\{  \dh  {\cal{P}}  , \dh  {\cal{L}}   \} _h = 
\sum_{ \Omega } 
\left( { \delta \dh {\cal{P}}  \over \delta { \bf u }  } \right)^T 
\dh{ {\cal{D}} } 
\left(
{ \delta \dh {\cal{L}}  \over \delta { \bf u }   } 
\right) h
\end{equation}
defines a Poisson bracket for functionals from $\dh {\cal F}$, 
i.e. the bracket $\{ \cdot , \cdot  \}_{h}$ is 
skew--symmetric and satisfies 
the Jacobi identity.   
Then we can choose some 
 approximation $ \dh{ {\cal{H}} } $ 
of the Hamiltonian $ {\cal{H}} $ to  obtain the set of 
semidiscrete Hamiltonian equations 
\begin{equation} \label{eq1} 
\dot{ \bf u } _j   = \dh{ {\cal{D}} } 
\left( { \delta \dh {\cal{H}}  \over \delta { \bf u }_j   } \right), 
\qquad j \in \mathbb{Z}
\end{equation}
which approximate equation (\ref{eqy2}) on mesh (\ref{meshom}).

For the space discretization of the canonical Hamiltonian PDEs 
we can keep the canonical operator $J$ since it generates 
a discrete Poisson bracket, namely  
$$
\{  \dh  {\cal{P}}  , \dh  {\cal{L}}   \} _h = 
\sum_{ \Omega } 
\sum_{i=1} ^n 
\left( 
 { \delta \dh {\cal{P}} \over \delta  v^i   }  
 { \delta \dh {\cal{L}} \over \delta  w^i   } 
- 
 { \delta \dh {\cal{P}} \over \delta  w^i   }  
 { \delta \dh {\cal{L}} \over \delta  v^i   } 
\right) h , 
$$
and take a discretization $\dh {\cal H } $ of the Hamiltonian functional 
$ {\cal H } $. This procedure provides us 
semidiscrete canonical Hamiltonian equations 
\begin{equation} \label{eqtr5} 
\dot{\bf v}_j = { \delta \dh {\cal H } \over \delta { { \bf w} _j} } , \qquad 
\dot{\bf w}_j = - { \delta \dh {\cal H } \over \delta { { \bf v} _j} }, 
\qquad  j \in \mathbb{Z} ,  
\end{equation}
where we use vector notations 
$  { \bf v} = ( v^1 , ..., v^n )^T$, $  { \bf w} = ( w^1 , ..., w^n )^T$.

\section{Symmetries and conservation laws}

\subsection{Invariance of semidiscrete equations}

Let $\dh Z$ be the space of sequences of variables 
$( t, x, h, { \bf u}  , \dh { \bf u} { } _1,  \dh { \bf u}  { }_2, ...)$ 
and $\dh {\cal A}$ be the 
space of analytic functions of a finite number of variables $z$ from 
$\dh Z$. 

Invariance of the semidiscrete equations 
\begin{equation}  \label{eqta1}
\dot{\bf u}  = {\bf F}  (z),  \quad { F}_i  \in \dh {\cal A} 
\end{equation}
defined in the points of some two--dimensional mesh $\Omega$ 
was considered in~\cite{[Kozl1]}. 
Symmetries of  equations (\ref{eqta1}) 
are transformations generated 
by vector fields of the form 
\begin{equation} \label{eqqqa} 
X =   \xi^t (z)   { \ddt} +  \xi^x (z)  { \ddx} + \eta^i  (z) {\ddui} 
+ \cdots , \qquad  \xi^t,  \xi^x,  \eta ^i \in \dh {\cal A } ,
\end{equation}
which leave the equations and the mesh invariant. 
The infinitesimal criterion of invariance 
can be presented by the three conditions 
\begin{equation}  \label{eqa9}
  X (  \dot{ \bf u} - { \bf F} (z) )   = 0 ; 
\end{equation}
\begin{equation}  \label{eqqs1}
 \df{ { {D}} }  \dd{ { {D}} } ( \xi^x)  = 0 ;  \qquad 
{ { {D}} }_{t}  ( \xi^x)  = 0  ; 
\end{equation}
which are to be satisfied on the solutions of (\ref{eqta1}). 
Condition (\ref{eqa9})  requires the invariance of Eqs.~(\ref{eqta1}) 
while conditions (\ref{eqqs1}) that of mesh $\Omega$. 
The operator $X$ must be 
prolonged on all variables appearing in the equation (\ref{eqta1})   
\begin{equation} \label{equca2} 
X =  
\xi^t   { \ddt} +  \xi^x  { \ddx} + \eta^i  {\ddui}
+ \phi^i  { \partial \over \partial \dot{u}^i  }
+ \dh  \zeta { } _1^i { \partial \over \partial \dh u { }_1 ^i  }  
+ \dh \zeta { } _2^i { \partial \over \partial \dh u { }_2 ^i } + \cdots  + 
\dd D ( \xi^x ) { \partial \over \partial h } . 
\end{equation}
On the uniform in space grid 
the coefficients of the prolonged operator are defined by the prolongation 
formulas: 
\begin{equation} \label{equca3} 
\begin{array}{c}
\phi^i = D_t ({ \eta ^i }) 
-  \dot{u}^i     D_t ( \xi^t ) -   u_1 ^i   D_t ( \xi^x ) , \quad 
\dh \zeta { } _1 ^i  = \dd D (\eta ^i ) 
- S_{+}  (\dot{u}^i )  {\dd D} (\xi^t) - \dh u { }_1 ^i  {\dd D} (\xi^x), \\
\\
\dh \zeta { }_2 ^i = {\df D}{ \dd D} ( \eta ^i  ) 
- 2 \ {\dh u { }_2 ^i  } \dd D (\xi^x) 
- { 1 \over h } { S_{+} } (\dot{u} ^i )  {\dd D} (\xi^t) 
+ { 1 \over h } { S_{-} }(\dot{u} ^i )  {\df D} (\xi^t) , \quad \cdots  \\
\end{array}
\end{equation}
Note that $u_1^i  = D_x (u^i ) $ is the ``continuous'' derivative. It is supposed 
to be in some discrete representation, for example, 
$ \dh { \tilde{D} } { }_0 (u)$, which will be introduced below.

Operators of the form (\ref{eqqqa}) are called Lie--B\"{a}cklund 
(or {\it generalized}) symmetries. 
It is a difficult task to find Lie--B\"{a}cklund operators admitted by 
discrete equations. However, Lie point symmetries 
\begin{equation}  \label{eqa8}
X  =  \xi^t (t,x, {\bf u} )  {\ddt} + \xi^x  (t,x, { \bf u} ) {\ddx} 
+  \eta^i   (t,x, { \bf u} ) {\ddui} + \cdots , 
\end{equation} 
where coefficients $\xi^t$,  $\xi^x$ and $\eta$ 
depend only on dependent and independent variables, 
are easier to detect since 
such symmetries are given by the same vector fields 
in the continuous and discrete cases~\cite{[5c]}. 
Practically, one can check 
whether Lie point symmetries admitted by the underlying PDEs are 
admitted by the semidiscrete equations or not. Although this 
procedure does not guarantee that we find all symmetries, 
it let us avoid solving discrete determining equations.

\subsection{Factorization of operators}

Following~\cite{[5c]},\cite{[5b]}, 
let us consider a special operation of left multiplication of 
a Lie-B\"{a}cklund operator by an analytic function 
$ \tilde{\xi} (z) \in  \dh {\cal A } $: 
\begin{equation} \label{cenkb}
\tilde{\xi} * X = 
\tilde{\xi} \xi^t  {\ddt} + 
\tilde{\xi} \xi^x  {\ddx} + 
\tilde{\xi} \eta ^i { \ddui } + \cdots +
\dd D ( \tilde{\xi} \xi^x ) { \partial \over \partial h } . 
\end{equation}
The first coordinates in operator $\tilde{\xi}  * X$ 
are multiplied by $ \tilde{\xi} (z)$ while the remaining coordinates 
are computed according to the prolongation formulas (\ref{equca3}).

The operator 
\begin{equation} \label{idea1}
\xi^t(z)* D_t + \xi^x (z)* { \dh D { } _x } ,
\end{equation}
where $D_t $ is the total time derivative operator and 
$\dh D { }_x $ is a discrete presentation 
of the total space derivative operator $D_x$, 
plays the role of an ideal of the Lie algebra of 
operators (\ref{eqqqa}) \cite{[Kozl1]}. 
There are  several possibilities to choose operator $ \dh D { }_x$. 
One can take a representation based on the 
right or  left discrete derivative~\cite{[5c]},\cite{[5b]}
\begin{equation} \label{eq13n}
\begin{array}{c}
{ D } ^+ ={\displaystyle  {\partial \over \partial x}} +
\tilde {\dd{D}}(u) {\displaystyle {\partial \over \partial u}} + \ldots,
\qquad 
{\displaystyle \tilde{\dd D}}={\displaystyle \sum_{n=1}^{\infty}
\frac{(-h)^{n-1}}{n} {\dd D}^n} , 
\\
\\
{ D} ^- = {\displaystyle \frac{\partial}{\partial x}} + \tilde{\df D}(u)
{\displaystyle \frac{\partial}{\partial u}} + \ldots, 
\qquad 
{\displaystyle \tilde{\df D}}={\displaystyle \sum_{n=1}^{\infty}
\frac{h^{n-1}}{n} {\df D}^n} ,  \\
\end{array}
\end{equation}
or the discrete representation based on 
the central difference derivative~\cite{[Kozl1]}
\begin{equation} \label{equt13n}
{ D } ^0 ={\displaystyle  {\partial \over \partial x}} +
\dh {\tilde{D}}  { } _0 (u) {\displaystyle {\partial \over \partial u}} + 
\cdots  , \qquad 
 \dh {\tilde{D}} { } _0 = 
\sum_{k = 0 } ^ { \infty}  
  \alpha_{2k+1}  h ^{2k}   \dh D { }_0 ^{2k+1}  , \qquad 
{ \dh {D}  { } _0 }   = { S_{+} - S_{-} \over 2h }  
\end{equation} 
with coefficients 
$$  
\alpha_{2k+1} = ( -1 ) ^ k { 1 \over 2 } 
{ 3 \over 4 } \cdots  {  2k -1 \over 2k  } 
 { 1 \over 2k+1  } =  { ( -1 ) ^ k \over 2k+1  } 
{ (2k-1)!! \over  2^k k!  } , 
\qquad 
(2k-1)!! = 1 \cdot 3  \cdot 5 \cdots (2k-1). 
$$

Let us mention that operator 
$ \dh {\tilde{D}} { }_  0 $ can be  presented in terms of powers 
of the shift operator $S_+$:  
\begin{equation} \label{eqtq1} 
 { \dh {\tilde{D}}  { } _0 }   = \sum_{k = - \infty} ^{ \infty}  c_k S_{k} , 
\qquad S_{k} = (S_+) ^k 
\end{equation}
with convergent coefficients $c_k $. 
To find this representation 
we rewrite this operator as 
\begin{equation} \label{eq14} 
\dh { \tilde{D}} { } _0 = 
{ 1 \over 2h } 
\sum_{k = 0 } ^{\infty} 
{  \alpha_{2k+1}    \over 2 ^{2k} }  ( S_{+} - S_{-} ) ^{2k+1} . 
\end{equation}
Using 
$$
( S_{+} - S_{-} )   ^{2k+1} = 
\sum_{ j = 0 } ^{ k } (-1) ^j 
{ ( 2k+1) !  \over j! ( 2k+1 - j ) ! } S _{2k+1 - 2j } + 
\sum_{ j = k + 1 } ^{ 2k+1 } (-1) ^j 
{ ( 2k+1) !  \over j! ( 2k+1 - j ) ! } S _{2k+1 - 2j }
$$
$$
= \sum_{ p = 0 } ^{ k } (-1) ^{k-p} 
{ ( 2k+1) !  \over (k-p) ! ( k+ 1 + p ) ! } S _{ 2p+1 } + 
\sum_{ p = 0 } ^{ k } (-1) ^{k+1+p} 
{ ( 2k+1) !  \over (k-p) ! ( k+ 1 + p ) ! } S _{- 2p-1 }  
$$
$$
= 
(-1)^k \sum_{ p = 0 } ^{ k } (-1) ^{p} 
{ ( 2k+1) !  \over (k-p) ! ( k+ 1 + p ) ! } 
( S _{ 2p+1 } - S _{- 2p- 1 }  ) , 
$$
we obtain 
\begin{equation} \label{eqrr15} 
\dh { \tilde{D}} { } _0 = 
 { 1 \over 2h } 
\sum_{ p = 0 } ^{ \infty } 
c_p ( S _{ 2p+1 } - S _{- 2p - 1 }  ) , 
\end{equation}
where the coefficients are given by the series 
\begin{equation} \label{eq15}
\begin{array}{ccl} 
c_p & =  & 
{ \displaystyle 
( -1 )^p \sum_{k = p } ^{\infty} 
 (-1)^k { \alpha_{2k+1}    \over 2 ^{2k} } 
{ ( 2k+1) !  \over (k-p) ! ( k+ 1 + p ) ! } } \\
 & & \\
& = & 
{ \displaystyle 
( -1 )^p \sum_{k = p } ^{\infty} 
 { 1 \over   k+ 1 + p }  { 1  \over 2 ^{2k} } 
 { (( 2k-1)!!)^2   \over (k-p) ! ( k + p ) ! } } 
\end{array}
\end{equation}

\begin{lemma}\label{lemma:2} The series (\ref{eq15}) defining coefficients 
$c_{p}$ for $p \in \mathbb{N}  $ are convergent. 
\end{lemma}

\noindent { \it Proof: } Using $(k+p)! (k-p)! \geq (k!)^2 $ we get

$$
| c_{p} | \leq  \sum _{k =p  } ^{ \infty }  
{ 1 \over  k+1+p  } 
\left(    
{ (2k) !   \over ( 2^{k} (k)! )^2   }
\right)^2 . 
$$
We use the bounds on the factorial provided by 
the Stirling's expansion
$$
n^n \exp( -n) \sqrt{ 2 \pi n } < 
n! < 
n^n  \sqrt{ 2 \pi n }\exp \left( -n + { 1 \over 12 n } \right) 
\quad \mbox{for} \quad n \in \mathbb{N} 
$$
to obtain the inequality 
$$
{ 1 \over  k+1+p  } 
\left(    
{ (2k) !   \over ( 2^{k} (k)! )^2   } \right)^2 < 
{ 1 \over \pi  k (k+1+p) }  \exp \left( { 1 \over 12k } \right)
\quad \mbox{for} \quad k \in \mathbb{N} 
$$
that insures us that the series defining $c_p$ converges, 
moreover $ c_{p}  \sim  1  /| p|  $ as $ p \rightarrow \infty$.  

\hfill $\Box$

One can easily check that 
the operators $ \dhp{\tilde{D}} $ and $  \dhm{\tilde{D}} $ 
can not be presented in the form 
(\ref{eqtq1}) with convergent coefficients.

\begin{lemma}\label{lemma:8} The operator $ { \dh {\tilde{D}} { } _0 } $ is 
skew--adjoint  , i.e. 
$$
\sum_{ i = - \infty } ^ {  \infty }  v_i { \dh {\tilde{D} } { }_0 }   u_i h
= - \sum_{ i = - \infty } ^ {  \infty }  u_i { \dh {\tilde{D}} { } _0 } v_i h
$$
for $ \{ u_i \}$,   $ \{ v_i \}$, $i \in \mathbb{Z} $ 
such that $ u_i  ,   v_i  \rightarrow 0 $ 
as $ i \rightarrow  \infty$ . 
\end{lemma}

\noindent { \it Proof: } The equality can be check with the 
help of (\ref{eqrr15}). 
\hfill $\Box$

\medskip


In what follows we will use $  { {D} }   ^0   $.  
With the help of the ideal $ \xi^t * D_t + \xi^x * { D^0 } $ 
we can find the evolutionary operator corresponding to 
the operator (\ref{eqqqa}) 
\begin{equation}  \label{wddqrr2}
\bar{X} = X - \xi^t * D_t + \xi^x * { D^0 } 
=    \bar{\eta} ^i   { \ddui } +  \cdots ,    \qquad  
\bar{\eta} ^i  = \eta ^i   - \xi^t \dot{ u} ^i - 
\xi ^x {  \dh {\tilde{D}} { } _0    }  ( u^i  ) 
\end{equation}
On the solutions of equations (\ref{eq1}) 
we can exclude time derivatives $\dot{ u} ^i$, $ i = 1,...,N$ 
from the coefficients of the evolutionary operator. 


It is important to note that although 
the factorization let us find evolutionary 
operators corresponding 
to given Lie point operators, generally, the obtained evolutionary 
operators does not have to be admitted by the semidiscrete equations 
which admit the original nonevolutionary operators. 
In the continuous case $ \xi^t * D_t =\xi^t \cdot  D_t $ and 
$ \xi^x * D_x = \xi^x \cdot D_x $, i.e., the operation~$*$ is 
equivalent to a left multiplication of the prolonged operator~\cite{[5c]}. 
The operators  $ D_t$ and $ D_x$ are admitted by all differential equations 
and, consequently, the operators  
$ \xi^t * D_t$ and $ \xi^x * D_x $ are also admitted by all equations. 
It follows that an evolutionary operator 
is admitted if it corresponds to an admitted  nonevolutionary 
operator~\cite{[2]}.

This result does not hold in the semidiscrete case. 
In the general case operator $ \xi^t * D_t = \xi^t \cdot  D_t$ 
is admitted, but operator $ \xi^x * { D }  ^0  $ is not. 
Thus, having obtained an evolutionary operator, 
one has to check that this operator or  
used in the factorization operator 
$ \xi^x * { D }  ^0  $ is admitted by 
the considered semidiscrete equations.

We can provide only a very restricted class of operators 
$ \xi^x * { D }  ^0  $ which are admitted by an arbitrary 
discrete equation.

\begin{lemma}\label{lemma:13} For $ \xi ^x  $ such that 
$  \dhp  D (  \xi^x ) = 0 $ we have 
$$
 \xi^x * { D }  ^0 =  \xi^x \cdot { D }  ^0
$$
and, consequently, this operator is admitted by all difference equations. 
\end{lemma}

\noindent { \it Proof: } The result follows from the prolongation 
formulas. 
\hfill $\Box$

\medskip

\noindent { \bf Remark} $\ $
The condition $  \dhp D (  \xi^x ) = 0 $ is very restrictive, 
but its multi--dimensional analog 
$$
  \dhpi  D  (  \xi^{x_i}  ) = 0 , 
$$ 
where $x_i$  is an independent  space variable and $\dhpi D$ is 
the right discrete derivative with respect to $x_i$, 
leaves more freedom. For example, it allows rotations 
in $X_i X_j$, $i \neq j$ planes 
$$
X_{i,j} = 
x_{i} { \partial \over \partial x_j } -  
x_{j} { \partial \over \partial x_i }. 
$$

\subsection{Conservation laws}

For a system of Hamiltonian equations (\ref{eqtr5})  
considered on the grid (\ref{meshom}) 
we have the following 
types of the conservation laws:

\medskip

\noindent {\bf 1.}  Conservation of symplecticity. 

\medskip

Due to their canonical form semidiscrete equations 
possess the conservation of symplecticity 
\begin{equation} \label{symp1} 
{ d  \over d t } { \dh { \bf \omega } } = 0 , \qquad 
 { \dh { \bf \omega } }  = 
\sum _{ j = - \infty } ^ { \infty }  
   d { \bf v} _j  \wedge d { \bf w} _j \  h
= \sum _{ j = - \infty } ^ { \infty }  
 \sum_{i =1} ^n  d v^i _j  \wedge d w^i _j \  h ,
\end{equation} 
where  $ d { \bf v} _j  = ( d  v^1 _j , ..., d  v^n _j) ^T  $ 
and $d { \bf w} _j =  ( d  w^1 _j , ..., d  w^n _j) ^T  $ are 
solutions of the variational equations 
$$
d \dot{v}^i_j   =  \sum_{k=1}^n \sum_{l}   
{ \partial \over \partial   w^k_{j+l} }     
\left(   { \delta \dh {\cal{H}}  \over \delta w^i_j  } \right)  d w^k_{j+l} +
 \sum_{k=1}^n \sum_{l}    
{ \partial \over \partial   v^k_{j+l} }     
\left(   { \delta \dh {\cal{H}}  \over \delta w^i_j  } \right)  d v^k_{j+l} ,
$$
$$
d \dot{w}^i_j   =  - \sum_{k=1}^n \sum_{l}   
{ \partial \over \partial   w^k_{j+l} }   
\left( { \delta \dh {\cal{H}}  \over \delta v^i_j    } \right)  d w^k_{j+l} -  
\sum_{k=1}^n \sum_{l}   
{ \partial \over \partial   v^k_{j+l} }   
\left( { \delta \dh {\cal{H}}  \over \delta v^i_j    } \right)  d v^k_{j+l} .  
$$
We suppose that  $ d { \bf v} _j , \  d { \bf w} _j \rightarrow 0 $ as 
$ j  \rightarrow \infty$. Differentiating the 2--form  
${ \dh { \bf \omega } }$ 
$$
{ d  \over d t } { \dh { \bf \omega } } = 
\sum _{ j = - \infty } ^ { \infty }  
 \sum_{i =1} ^n  d \dot{v}^i _j  \wedge d w^i _j \  h  + 
\sum _{ j = - \infty } ^ { \infty }  
 \sum_{i =1} ^n  d v^i _j  \wedge d \dot{w}^i _j \  h   
$$
$$
= \sum _{ j, l = - \infty } ^ { \infty } \sum_{i,k =1} ^n 
{ \partial \over \partial   w^k_{j+l} }     
\left(   { \delta \dh {\cal{H}}  \over \delta w^i_j    } \right)  d w^k_{j+l}
\wedge d w^i _j \  h + 
\sum _{ j,l  = - \infty } ^ { \infty } \sum_{i,k =1} ^n 
 { \partial \over \partial   v^k_{j+l} }     
\left(   { \delta \dh {\cal{H}}  \over \delta w^i_j    } \right)  d v^k_{j+l} 
\wedge d w^i _j \  h 
$$
$$ 
- 
\sum _{ j, l = - \infty } ^ { \infty } \sum_{i,k =1} ^n 
{ \partial \over \partial   w^k_{j+l} }   
\left( { \delta \dh {\cal{H}}  \over \delta v^i_j    } \right)  
d v^i _j  \wedge  d w^k_{j+l} \ h
- 
\sum _{ j, l = - \infty } ^ { \infty } \sum_{i,k =1} ^n 
{ \partial \over \partial   v^k_{j+l} }   
\left( { \delta \dh {\cal{H}}  \over \delta v^i_j    } \right)   
d v^i _j  \wedge  d v^k_{j+l} \ h 
$$
and using 
$$
{ \partial \over \partial   w^k_{j+l}  }     
\left(   { \delta \dh {\cal{H}}  \over \delta w^i_j    } \right) = 
{ \partial \over \partial   w^k_{j}   }   
\left(   { \delta \dh {\cal{H}}  \over \delta w^i_{j+l}    } \right), 
\qquad
{ \partial \over \partial   v^k_{j+l}  }    
\left(   { \delta \dh {\cal{H}}  \over \delta v^i_j    } \right) = 
{ \partial \over \partial   v^k_{j}   }   
\left(   { \delta \dh {\cal{H}}  \over \delta v^i_{j+l}    } \right),
$$
$$
{ \partial \over \partial   v^k_{j+l}  }    
\left(   { \delta \dh {\cal{H}}  \over \delta w^i_j    } \right) = 
{ \partial \over \partial   w^k_{j}   }   
\left(   { \delta \dh {\cal{H}}  \over \delta v^i_{j+l}    } \right), 
$$
we obtain the conservation of the symplectic form.

This is a generalization of the symplectic structure for the Hamiltonian 
ODEs~\cite{[Sanz]} to the infinite set 
of semidiscrete equations. 
Let us note that recently proposed 
multi--symplectic formulation of PDEs \cite{[Brid1]}--\cite{[Reich3]} 
allows to consider local conservation of symplecticity, 
which we do not have in the present framework.

Underlying equations (\ref{equ5}) possess  
the conservation of symplecticity 
\begin{equation} \label{symp2} 
{ d  \over d t } { \bf \omega }  = 0 , \qquad 
 { \bf \omega } = \int  d {\bf v } \wedge d { \bf w}  \  d x =  
\int \sum_{i = 1 }^n  d v^i \wedge d w^i  \ d x  
\end{equation} 
that is the continuous limit of (\ref{symp1}). In this case 
$  d { \bf v} $ and $  d { \bf w} $ are solutions of the 
variational equations for  (\ref{equ5}) 
$$   
d {v}^i_t   = 
\hat{d} \left(   { \delta {\cal{H}}  \over \delta w^i    } \right) = 
\sum_{k=1}^n  \sum_{l} 
 { \partial \over \partial   w^k_l }     
\left(   { \delta  {\cal{H}}  \over \delta w^i    } \right)   d w^k_l +
\sum_{k=1}^n  \sum_{l} 
{ \partial \over \partial   v^k_l }     
\left(   { \delta  {\cal{H}}  \over \delta w^i    } \right)   d v^k_l ;
$$
$$
d {w}^i_t   =  
- \hat{d} \left( { \delta  {\cal{H}}  \over \delta v^i    } \right) = 
- \sum_{k=1}^n  \sum_{l} 
{ \partial \over \partial   w^k_l }   
\left( { \delta  {\cal{H}}  \over \delta v^i    } \right)  d w^k_l 
- \sum_{k=1}^n  \sum_{l} 
{ \partial \over \partial   v^k_l }   
\left( { \delta  {\cal{H}}  \over \delta v^i    } \right)  d v^k_l   ;  
$$
where the operator $\hat{d}$ denotes the vertical differential of the 
vertical form it acts on. 
In our case we have vertical 0--forms, i.e. 
functions. We assume that $  d { \bf v} $, $  d { \bf w} $ and 
their space derivatives 
which appear in the variational equations 
decay at $ x \rightarrow \infty$. Let us 
show that the conservation of symplecticity using the variational complex 
(see, for example, \cite{[2]}):
$$
{ d  \over d t } { \bf \omega }  = 
\int \left( \sum_{i = 1 }^n  d v^i_t \wedge d w^i  + 
            \sum_{i = 1 }^n  d v^i \wedge d w^i_t  \right)  \ d x 
$$
$$
= \int \left(  
\sum_{i = 1 }^n  
\hat{d} \left(   { \delta  {\cal{H}}  \over \delta w^i    } \right)   
\wedge d w^i - 
\sum_{i = 1 }^n  d v^i \wedge 
\hat{d} \left( { \delta  {\cal{H}}  \over \delta v^i    } \right) 
\right)  \ d x 
$$
$$
= \int \hat{d} 
\left( \sum_{i = 1 }^n   
{ \delta  {\cal{H}}  \over \delta v^i    }  d v^i + 
{ \delta  {\cal{H}}  \over \delta w^i    }  d w^i 
\right) \ dx 
= \delta 
\int 
\left( \sum_{i = 1 }^n   
{ \delta  {\cal{H}}  \over \delta v^i    }  d v^i + 
{ \delta  {\cal{H}}  \over \delta w^i    }  d w^i 
\right) \ dx
$$
Using the integration by part,  we obtain that 
$$
\int 
\sum_{i = 1 }^n   
{ \delta  {\cal{H}}  \over \delta v^i    }  d v^i \ dx 
= 
\int 
\sum_{i = 1 }^n  
\sum_{k} ( -D_x) ^k 
{ \partial   H  \over \partial  v^i_k   }  d v^i  \ dx 
= \int 
\sum_{i = 1 }^n  
\sum_{k}  
{ \partial   H  \over \partial  v^i_k   }  d v^i_k  \ dx 
$$
because $d v^i_k$,  $i = 1, ..., n$ tend to zero as $x \rightarrow \infty$. 
Since the same result is valid for variables ${\bf w}$ we get 
$$ 
\int 
\left( \sum_{i = 1 }^n   
{ \delta  {\cal{H}}  \over \delta v^i    }  d v^i + 
{ \delta  {\cal{H}}  \over \delta w^i    }  d w^i 
\right) \ dx 
= \int \hat{d}  H  \ dx 
= \delta {\cal{H}} 
$$
so that 
$$
{ d  \over d t } { \bf \omega }  = \delta ^2   {\cal{H}} = 0  
$$
as follows from the exactness of the variational complex.

\medskip

\noindent {\bf 2.} Conservation of distinguished functionals 

\medskip

\noindent {\bf Definition} For a given Hamiltonian operator $\dh {\cal D}$ 
a distinguished functional is a functional 
$\dh {\cal C} (x,h, {\dh { \bf u} }  ^{(n)} )  $ such that 
\begin{equation} \label{eqy13} 
\dh{ {\cal{D}} } 
\left( { \delta \dh {\cal{C}}  \over \delta { \bf u }  } \right) = 0 
\end{equation} 
It follows that 
a functional is distinguished if and only if its Poisson bracket with 
every other functional is trivial: 
\begin{equation} \label{eqy14} 
\{ \dh {\cal C} , \dh  {\cal H} \}_h  = 0 
\quad \mbox{for any } \quad  \dh {\cal H} \in \dh {\cal F} . 
\end{equation}

Distinguished functionals are conserved by semidiscrete equations 
originating from any Hamiltonian functional. 
For the canonical bracket 
there are no nontrivial distinguished functionals
so that we do not get conservation laws of this type. 

\medskip

\noindent {\bf 3.} Hamiltonian form of Noether's theorem. 

\medskip

\noindent {\bf Definition} The Hamiltonian vector field 
associated with a functional $\dh {\cal P}$ 
is the unique 
smooth vector field $X_P$ satisfying 
\begin{equation} \label{equa12} 
 X_P ( \dh {\cal F}  ) = \{ \dh {\cal F}, \dh {\cal P} \}_h 
\end{equation}
In the coordinate form it can be presented as the operator 
\begin{equation} \label{equ12} 
X_{P} = \dh {\cal D} 
\left( { \delta \dh {\cal{P}} \over \delta u^i   } \right) { \ddui} 
\end{equation}

Some symmetries (\ref{eqqqa}) are given as Hamiltonian vector fields or are 
equivalent to Hamiltonian vector fields under factorization~(\ref{wddqrr2}). 
Such vector fields let us use the following theorem~\cite{[Kozl1]}.

\begin{theorem} \label{ther3:2} For a Hamiltonian system of 
semidiscrete evolution equations (\ref{eq1}) a Hamiltonian vector field 
${X}_P$ determines a generalized symmetry of the system if and only 
if there is an equivalent functional 
$ \dh {\tilde{ \cal P}} =  \dh { \cal P } - \dh { \cal C } $, 
differing from $\dh {\cal P} $ only by  a time--dependent 
distinguished functional $ \dh { \cal C } (t,x,h,u^{(n)} ) $, such that 
 $ \dh { \tilde{ \cal P} } $ determines a conservation law. 
\end{theorem}

For the operator $J$ generating the 
canonical bracket 
a Hamiltonian vector field has the form 
\begin{equation} \label{symm1} 
X_{P} =  { \delta \dh {\cal{P}}   \over \delta w ^i } 
{ \partial  \over  \partial v^i }  - 
 { \delta \dh {\cal{P}}   \over \delta v ^i } 
{ \partial  \over  \partial w^i }, 
\end{equation}
where $ \dh { \cal P} $ is the generating functional. 
The canonical bracket has only trivial time--dependent 
functionals, i.e. functions $f(t)$. 
Thus, for a Hamiltonian symmetry 
(\ref{symm1}) there corresponds a conservation law 
$ \dh {\tilde{ \cal P}} =  \dh { \cal P } - f(t) $, where the 
function $  f(t) $ needs to be found with the help of the considered 
equations. In a particular case when 
the Hamiltonian functional and the considered Hamiltonian symmetry are 
time independent we obtain a linear function $ f(t) = a t + b$, 
$a,b = const$.

\section{The nonlinear wave equation}

In this section we consider 
the nonlinear wave equation 
\begin{equation} \label{syst2} 
v_{tt} =  v _{xx} - V'(v) , 
\end{equation}
where $V(v)$ is some smooth function. 
For simplicity, we consider the case of scalar $v$. 
With the help of a new variable $ w = v_t$, the equation (\ref{syst2}) 
can be rewritten as the system 
\begin{equation} \label{sys2} 
\left\{ 
\begin{array}{l} 
{ \displaystyle 
{ v_t } = w  }; \\
\\
{ \displaystyle 
{ w_t } = v_{xx} - V'(v)  }. \\
\end{array}
\right.
\end{equation} 
This is a canonical Hamiltonian system generated by 
the Hamiltonian functional 
\begin{equation}  \label{swe45}
{ \cal H }  = \int \left( { w^2 \over 2 } + { v^2 _x \over 2 }  
+ V(v) \right) dx. 
\end{equation} 
Let us take the following approximation of the Hamiltonian functional 
\begin{equation} \label{ham21} 
{\cal{H}} _{h}  = 
\sum_{ \Omega } H [v,w] h , \qquad 
H =  { w ^2 \over 2 }   + { { \dh v { }_1 ^2 } \over 2 }  + V(v) . 
\end{equation} 
It provides us the system of the semidiscrete equations
\begin{equation} \label{sys5} 
\left\{ 
\begin{array}{l} 
{ \displaystyle 
{ \dot{v} } =  w   }; \\
\\
{ \displaystyle 
{ \dot{w} } 
=  { \dh v { }_2 }  - V' (v)  } . \\
\end{array}
\right.
\end{equation}

For arbitrary $V(v)$ the admitted transformation group for 
(\ref{sys5}) is 
two--dimensional. Its Lie algebra is spanned by the 
operators 
\begin{equation} \label{sym1} 
X_{1} = {\ddt} , \qquad  X_2 = {\ddx} . 
\end{equation} 
Lorentz transformation 
$$
X_{3} =  x {\ddt} + t {\ddx}, 
$$
admitted by underlying system (\ref{sys2}) 
with arbitrary $V(v)$~\cite{[72]}, is lost under discretization 
since it brakes the mesh invariance. 
Thus in the case of arbitrary $V(v)$ Noether's theorem gives 
two conservation laws:

\medskip
 
\noindent {  \bf 1.}  The time translation $X_1$ 
leads to the conservation of the Hamiltonian 
functional $ \dh { \cal H} $. Indeed, the factorization 
of operator $X_1$ gives the evolutionary vector field 
$$
\bar{X}_{1}  = w  { \partial \over \partial v } + 
\left( { \dh v { }_2 } - V' (v) \right) 
  { \partial \over \partial w }, 
$$
which is generated by the Hamiltonian functional. 
Thus we obtain conservation 
of Hamiltonian functional (\ref{ham21}) that stands for 
the conservation of energy.  

\medskip

\noindent  { \bf 2.}  The space translation $X_2$ corresponds 
to the evolutionary operator 
\begin{equation} \label{eq11} 
\bar{X}_2  = 
\dh { \tilde{D}} { }_0 ( v ) { \ddv} + 
\dh { \tilde{D}} { }_0 ( w ) { \ddw} .  
\end{equation}
Checking condition (\ref{symm1}) for the coefficients of operator 
$\bar{X} _2 $, we find the generating functional 
\begin{equation} \label{eq17} 
\dh { \cal P} { }_2  = \sum _{\Omega} 
w  { \dh { \tilde{D} } { }_0 } ( v )  h 
=
-  \sum _{\Omega } 
v  { \dh { \tilde{D} } { } _0  } ( w )  h . 
\end{equation}
This functional is a conservation law of the semidiscrete equations 
(\ref{sys5}). 
In the continuous limit it corresponds to the functional 
\begin{equation} \label{eq19} 
{\cal P}_2   =
\int  w v _x  dx = - \int v w _x  dx, 
\end{equation}
which is a conservation law of the Eqs.~(\ref{sys2}). 
In physical applications  
this conservation law is referred as linear momentum.

 In order to obtain this conservation law 
we need to consider ``nonlocal'' functionals, 
i.e.  functionals which are  defined as double sums  over the mesh points, 
since operator ${ \dh { \tilde{D} } { }_0  }  $ is a sum over 
infinitely many mesh points.

\medskip

Let us consider the special cases of the potential $ V(v)$ 
which lead to additional Hamiltonian symmetries 

 \medskip

\noindent {  \bf a)} For the quadratic potential  
$ V(v) = C { v ^2 \over 2 } $ 
there is an infinite series of conservation laws for the wave 
system (\ref{sys5}). In this case 
the semidiscrete system 
admits the infinite set of Hamiltonian operators
$$
Y_k =  { \dh D { }_0}  ( \dh v { }_{2k} ) 
{ \partial \over \partial v } +   
{ \dh D { }_0}   ( \dh w { }_{2k} ) 
{ \partial \over \partial w }, \qquad 
k = 0,1,... , 
$$
which provides  us 
the infinite set of the conserved functionals 
$$
{ \dh R { }_{k} } = 
\sum_{ \Omega } 
w { \dh D { }_0}  ( \dh v { }_{2k} )  h = 
-  \sum_{\Omega}     
v { \dh D { }_0}  ( \dh w { }_{2k} )  h
$$
In the continuous limit these conservation laws correspond to the 
functionals 
$$
 {  R_{k} } = \int   v w_{(2k+1)} dx, 
$$
which are conserved quantities for the system (\ref{sys2}).

Since for the quadratic potential we get the system of linear 
equations their solutions possess a superposition principle. It is 
reflected in the invariance with respect to the symmetry 
$$
Z =  \alpha ( t , x)   
{ \partial \over \partial v } + 
 \alpha _t ( t , x) 
 { \partial \over \partial w }, 
$$
where the function $ \alpha ( t , x) $ is an arbitrary solution 
of the equation 
\begin{equation}  \label{cond4} 
\alpha_{tt} ( t , x )  = 
{  \alpha (  t, x + h  ) - 2 \alpha (  t, x ) + \alpha (  t , x - h  ) 
\over h^2  }  - C \alpha (  t, x ) . 
\end{equation} 
The operator $Z$ is Hamiltonian. It  corresponds to the 
functional
$$
{ \dh T } = 
\sum_{ \Omega } 
( \alpha ( t , x )  w  - \alpha_t  ( t , x )  v )  h , 
$$
which is a  conservation law of the semidiscrete linear system 
if function $ \alpha ( t , x) $ satisfies  
the equation (\ref{cond4}).  In the continuous limit 
this functional goes into the functional
$$
{  T  } = \int 
 (  \alpha  ( t , x )  w  - \alpha _t ( t , x )  v ) dx  , 
$$
which is  a  conservation law of (\ref{sys2}) if the function 
$ \alpha (t , x)  $ satisfies the equation 
\begin{equation}  \label{tete4} 
\alpha_{tt} ( t , x )  = \alpha_{xx} ( t , x ) - C \alpha (  t, x ) . 
\end{equation}

\medskip 

\noindent {  \bf b)} Let us consider the case $C=0$, 
i.e. the wave system (\ref{sys5}) 
without a potential $V'(v) \equiv  0 $, in details. 
The symmetry $Z$ is 
specified by a solution  of the equation 
\begin{equation}  \label{cond5}
\alpha_{tt} ( t , x )  = 
{  \alpha (  t, x + h  ) - 2 \alpha (  t, x ) + \alpha (  t , x - h  ) 
\over h^2  } .
\end{equation} 
We present a number of the 
conservation laws corresponding the symmetry  $ Z$ 
taking particular solutions $ \alpha (t,x)  $ 
of the equation (\ref{cond5}) in Table~2.

Let us remark that the continuous conservation laws 
in the case $V'(v) \equiv  0 $ are given by the 
function $ \alpha(t,x) $ satisfying the equation (\ref{tete4}) 
with $ C = 0 $. The general  solution can be written down as 
$$
 \alpha(t,x) = \alpha_1(t - x) + \alpha _2(t +x), 
$$
where $\alpha_1$  and $\alpha_2$ are arbitrary functions. 

\bigskip

\begin{example}\label{ex1:1} As we mentioned before 
the factorized operator may fail to be admitted. 
Let us consider $V'(v) \equiv  0 $. In this case 
system (\ref{sys5}) admits the scaling symmetry  
 $$
 X_* = - t { \ddt} -  x { \ddx } +  w { \partial \over \partial w } . 
$$
The corresponding evolutionary vector field 
$$
\bar {X}_* = 
( t w + x { \dh  {\tilde{D}} { }_0  } (v) ) { \ddv } + 
(  w + t { \dh v { }_{2} }   + x { \dh  {\tilde{D}} { }_0  } (w) ) { \ddw } , 
$$
found with the help of factorization formula (\ref{wddqrr2}), 
is Hamiltonian. It is generated by the functional 
$$
{ \dh { \cal P} { }_* }  = t \ { \dh { \cal H}  } + 
\sum _{\Omega} 
x w { \dh  {\tilde{D}} { } _0  } (v) h.
$$
However, neither the functional ${ \dh { \cal P} { } _* } $ is 
a conservation law nor the canonical operator 
$ \bar {X}_* $ is a symmetry of the semidiscrete system 
(\ref{sys5}) with $V'(v) \equiv  0 $. It happens because 
the operator 
$$ 
x * { D }^0 =   
x  { \ddx} + x { \dh  {\tilde{D}} { }_0  } (v)  { \ddv } + 
x { \dh  {\tilde{D}} { }_0  } (w)  { \ddw }  + ...  ,
$$
which is used for the factorization
$$
\bar {X}_* = {X}_* - t * D_t - x * { D } ^0 , 
$$
is not admitted by the semidiscrete system:
the second equation does not allow this operator 
since
\begin{equation} \label{symtr} 
 2 {  \dh v { }_{2}  }  + 
x { \dh  {\tilde{D}} { }_0  } ( \dot{w} ) \neq  { \dhm  D }   { \dhp  D } 
( x  { \dh  {\tilde{D}} { }_0  } ( v )  )   
\end{equation}
on the solutions of the semidiscrete equations. 
In the limit $ h \rightarrow 0 $ (\ref{symtr}) 
turns into an equality 
and the continuous limit of ${ \dh { \cal P} { }_* }$, 
namely the functional 
$$
{ { \cal P} _* }  = 
t \ { \cal H } + \int  x w v_x  dx , 
$$
is a conservation law of the system (\ref{sys2}) with $ V' \equiv  0 $.

\end{example}

\section{The nonlinear Schr\"odinger equation } 
 
Another equation which can be cast into the 
canonical Hamiltonian form is the 
nonlinear Schr\"odinger equation (\ref{EQ2}) 
which arises in nonlinear optics. It describes the main features 
of the beam interaction with a nonlinear medium and 
is considered as the 
{\it basic equation of the nonlinear optics}~\cite{[71]}. 
The equation also has important applications in plasma physics~\cite{[Zah]}.

Let us consider the case of one--dimensional space 
\begin{equation}  \label{equa1} 
i \psi _t + \psi_{xx} + F' ( |\psi|^2 ) \psi = 0 . 
\end{equation}
For real and imaginary components $v$ and $w$ ($\psi = v + i w $) 
the Schr\"odinger equation can be rewritten 
as the system
\begin{equation} \label{eqte6} 
\left\{ 
\begin{array}{l} 
   v_t = - w _{xx} - F'( v^2 + w^2 )  w ;\\
\\
  w_t =  v _{xx} + F'( v^2 + w^2 ) v ; \\
\end{array}
\right.
\end{equation}
which is a canonical Hamiltonian system. 
It is generated by the Hamiltonian functional 
\begin{equation} \label{equ3} 
{ \cal H } = { 1 \over 2 } \int 
\left( |   \psi_x |^2   -  F( |\psi|^2 ) \right) dx . 
\end{equation}

The system of equations (\ref{eqte6}) in the case 
of arbitrary $F$ admits a four--dimensional transformation group 
presented by the operators~\cite{[71]}:  
\begin{equation} \label{eq7} 
\begin{array}{c}
{ \displaystyle 
X_1 = {\ddt} , \quad 
X_2 = {\ddx} , \quad 
X_3 = w { \ddv}  -   v { \ddw }  , \quad 
X_4 = 2t { \ddx}  - x w { \ddv} + x v { \ddw }  . } 
\end{array}
\end{equation}
In the case ${  F(z) = C { z ^2 \over 2}  } $ there is an 
additional scaling symmetry 
\begin{equation} \label{eq8} 
X_5 = 2t { \ddt} + x { \ddx}  - v { \ddv} - w  { \ddw }. 
\end{equation}
Physically, this particular case of $F(z)$ corresponds to an 
isotropic medium 
with the cubic polarizability 
(first approximation of the nonlinear polarizability).

Let us discretize the   Hamiltonian functional as 
\begin{equation} \label{eq9} 
\dh {\cal  H } = \sum_{ \Omega} H[v,w] h , \qquad 
H = { 1 \over 2} \left( 
\dh v { }_1 ^2 +  \dh w { }_1 ^2 -  F (   v ^2 +   w ^2 ) \right) , 
\end{equation}
then we obtain the 
system of semidiscrete equations 
\begin{equation} \label{eq10} 
\left\{
\begin{array}{l}
{ \displaystyle 
\dot{v}  = - \dh w { }_2 -  
F' (  v ^2 +   w  ^2 )  w } ; \\
\\
{ \displaystyle 
\dot{w}   =  \dh v { }_2 + 
F' (  v  ^2 +   w ^2 )  v  } .  \\
\end{array}
\right. 
\end{equation}

Now we are in a position to go through the symmetries (\ref{eq7}), (\ref{eq8}) 
to check whether they 
are preserved under the space discretization. 
If they are admitted by the semidiscrete system 
(\ref{eq10}), they may 
provide conservation laws according to Theorem~\ref{ther3:2}.

\medskip

\noindent { \bf 1.}  The time translation $X_1$ 
is admitted by system~(\ref{eq10}).  
The corresponding to $X_1$ evolutionary 
symmetry is generated by 
the functional~(\ref{eq9}). Thus we find that 
the Hamiltonian functional is a conservation law 
of Eqs.~(\ref{eq10}). Its 
continuous limit is the functional $ { \cal  H } $. 
Physically, we interpret this conservation law as conservation of energy.

\medskip

\noindent {  \bf 2.} The space translation $X_2$ 
corresponds to the evolutionary operator (\ref{eq11}), which 
we already examined. 
This  symmetry  leads us to the conservation law
$ \dh {\cal{P}} { } _ 2 $ (see (\ref{eq17})).

\medskip

\noindent { \bf 3. } The evolutionary symmetry  $  X_{3} $ 
is generated by the functional 
\begin{equation} \label{eq20} 
\dh {\cal P} { }_3  = { 1\over 2 } \sum_{ \Omega }  
 \left( v ^2 + w ^2 \right) h , 
\end{equation}
which is a conservation of mass for the semidiscrete system 
(\ref{eq10}). 
In the continuous limit it goes into the functional  
\begin{equation} \label{eq21} 
{\cal P} _3  = { 1\over 2 } \int 
 \left( v^2 + w^2 \right) dx . 
\end{equation}

\medskip

\noindent { \bf 4. }  Galilean transformation $X_4$ is not admitted 
in the semidiscrete case. Its coefficients violate the second 
mesh invariance condition in (\ref{eqqs1}). 
The action of the generated 
by $X_4$ transformation on the independent variables  
$$
\begin{array}{l}
\hat{t} = t ;  \qquad \hat{x} = x + 2t a 
\end{array}
$$
clearly shows that it destroys the mesh geometry. 
It follows that the continuous conservation law 
corresponding to the movement of the center of mass
\begin{equation} \label{eqt20} 
{\cal P} _4  = \int 
\left( { x \over 2 } ( v^2 + w^2 ) + t ( w v _x -  v w_x ) 
\right) dx
\end{equation}
has no counterpart in the semidiscrete case. 

\bigskip
\noindent { \bf 5. } The additional symmetry $X_5$ is admitted  by 
equations (\ref{eq10}) with quadratic $F$, but it is not Hamiltonian 
(both in the continuous and discrete cases). 

\bigskip

We sup up the considered symmetries and their generating functionals 
in the Table~2.

\section{Conclusions}

In the paper we considered semidiscrete canonical Hamiltonian equations 
and showed how to find conservation laws of such equations using 
Noether's theorem. 
Our interest to canonical Hamiltonian equations is also motivated 
by their connection with Euler--Lagrange equations~\cite{[Marsd]}.  

Many PDEs can be presented as Euler--Lagrange equations 
for appropriate  Lagrangian functionals~\cite{[2]},\cite{[10]} 
and can be rewritten as 
canonical Hamiltonian equations (\ref{equ5}). Let us 
illustrate this on the example of the nonlinear wave equation (\ref{syst2}) 
which  
is the Euler--Lagrange equation
\begin{equation} \label{pyoo1} 
{ \delta L \over \delta v } = 0 , \qquad 
{ \delta \over \delta v } = 
{ \partial \over \partial v } 
-  D_t { \partial \over \partial v_t } 
-  D_x { \partial \over \partial v_x } 
\end{equation} 
for the Lagrangian functional 
\begin{equation} \label{pyy1} 
{ \cal L } = \int \int    L ( v,v_t, v_x) dx dt , \qquad
   L =  { v_t ^2 \over 2 } -  { v_x ^2 \over 2 } - V(v).  
\end{equation} 
With the help of the Legendge transformation 
$$
w = { \partial L \over \partial v_t } = v_t , \qquad 
{ \cal H } = 
\int \left(   { \partial L \over \partial v_t } v_t - L \right) dx 
$$
we obtain the Hamiltonian functional (\ref{swe45}). 
It generates equations (\ref{sys2}), which are equivalent to (\ref{syst2}).

Similar connection between Euler--Lagrange equations 
and canonical Hamiltonian equations can be established 
in the semidiscrete case. Let us consider 
the semidiscretization of functional (\ref{pyy1}) 
$$
 \dh { \cal L } = \int \sum_{\Omega}   
{  \dh L }  ( v, \dot{v}, {\dh v { } _1}  ) h dt , \qquad
   { \dh L }  =  { \dot{v} ^2 \over 2 } -  { {\dh v { } _1 ^2 } \over 2 } - V(v). 
$$
Its Euler--Lagrange equation
\begin{equation} \label{prr1} 
{ \delta { \dh { \cal L } }   \over \delta v } = 0 , \qquad 
{ \delta \over \delta v } = 
{ \partial \over \partial v } 
-  D_t { \partial \over \partial \dot{v} } 
-  { \dhm D } { \partial \over \partial { \dh v { }_1 }  }  
\end{equation} 
has the form 
\begin{equation} \label{ptp3} 
\ddot{v} = \dh v { }_2 - V'(v) . 
\end{equation} 
We can introduce the discrete Hamiltonian functional 
$$
w = { \partial \dh L \over \partial \dot{v} } = \dot{v} , \qquad 
\dh { \cal H } = \sum_{\Omega}  
\left (   { \partial L \over \partial \dot{v} } \dot{v} - \dh L \right) h  , 
$$
which in our case is the functional (\ref{ham21}). It generates 
the semidiscrete evolution equations (\ref{sys5}), 
which are equivalent to (\ref{ptp3}).

Thus,  looking for conservation laws of the 
semidiscrete Euler--Lagrange equations,  
one can consider the equivalent canonical Hamiltonian systems 
and find conservation laws in the Hamiltonian framework. 
They have the form 
\begin{equation} \label{prr4} 
\int  { \cal P  } dx  \qquad \mbox{and} \qquad 
\sum _{\Omega} \dh  { \cal P  } h
\end{equation} 
in the continuous and semidiscrete cases correspondingly. 
If necessary,  the 
conservation laws can be rewritten in term 
of the original variables. 
For example,  the conservation laws of the semidiscrete wave 
system (\ref{sys5}) with an arbitrary potential $V(v)$  found in 
Section~4 can be rewritten in terms of variables used in 
the  Lagrangian approach
$$
\dh { \cal P } { }_1  = 
\sum_{\Omega}  
\left (  
{ \dot{v}^2 \over 2 } +  { { \dh v { }_1 ^2}  \over 2 } + V(v)
\right) h , \qquad 
\dh { \cal P } { }_2  = 
\sum_{\Omega}  
 \dot{v}  { \dh { \tilde{D} } { }_0 } ( v )  h . 
$$
For a quadratic potential we also have the infinite series of the 
conservation laws
$$
{ \dh R { }_{k} } = 
\sum_{ \Omega } 
\dot{v}  { \dh D { } _0}  ( \dh v { }_{2k} )  h = 
-  \sum_{\Omega}     
v { \dh D { }_0}  ( \dh {\dot{v}} { }_{2k} )  h , \qquad k = 0,1,2,...
$$  
and the conservation laws 
$$
{ \dh T  } = 
\sum_{ \Omega } 
( \alpha ( t , x )  \dot{v}  - \alpha _t ( t ,  x )  v )  h , 
$$
where the function  $\alpha ( t , x )  $ 
must satisfy the equation (\ref{cond4}).

One of the advantages of Lagrangian approach over Hamiltonian 
one is a possibility to find local conservation. 
In the Hamiltonian framework we can find only global conservation laws 
of the form (\ref{prr4}). 
This advantage of the Lagrangian approach is difficult to 
represent for the semidiscrete equations because 
as we have seen some conservation laws 
have densities which involve discrete presentations of 
the continuous derivatives 
and, consequently, are not local.

For simplicity we considered only the case of one--dimensional  space. 
The extension to the  multi--dimensional space   
is straightforward. 


\bigskip

\noindent {\bf \large Acknowledgments}  

\medskip

The author would like to thank Prof. C.~J.~Budd 
for stimulating discussions on Hamiltonian equations, 
in particular for pointing out that some semidiscrete equations 
can be cast into the Hamiltonian form. 
The research was sponsored in part 
by The Norwegian Research Council under contract 
no.~111038/410, through the SYNODE project. 
 

\eject

\noindent {\large  \bf Appendix}

\hoffset -2.5 truecm

\begin{center}
{ \bf Table~1.} A number of conservation laws for the wave system (4.5) 
without a potential $V'(v) \equiv  0 $, which 
correspond to particular cases of the symmetry 
$$
Z =  \alpha ( t , x)   
{ \partial \over \partial v } + 
 \alpha _t ( t , x) 
 { \partial \over \partial w }. 
$$
\end{center}

$$
\begin{array}[h]{|l|l|l|l|}
\hline
{\mbox{Function}}  \  \alpha(t,x) & 
\mbox{Operator} \  Z & \mbox{Discrete conservation law}  & 
\mbox{Continuous conservation law} \\ 
\hline
& &  & \\
\quad  1  & 
{ \displaystyle
{ \partial \over \partial v }   }  & 
{ \displaystyle
\sum_{ \Omega }   w   h  } & 
 {  \displaystyle \int   w   dx   } \\
& & & \\
\hline
& & &  \\
\quad   t  & 
{ \displaystyle 
   t   { \partial \over \partial v } + 
   { \partial \over \partial w } }  &  
{ \displaystyle  
\sum_{ \Omega }   ( t w - v )   h    } & 
{ \displaystyle  
\int    ( t w - v )   dx    } \\
& & &  \\
\hline
& & &  \\
\quad   x  & 
{ \displaystyle  
  x   { \partial \over \partial v }  } &  
{ \displaystyle  
\sum_{ \Omega }   x w    h   } &  
{ \displaystyle  
\int  x w    dx   } \\
& & & \\
\hline
& & & \\
\quad   tx  & 
{ \displaystyle 
   tx   { \partial \over \partial v } + 
  x  { \partial \over \partial w }  }  & 
{ \displaystyle 
 \sum_{ \Omega }   ( tx w - x v )   h  }  & 
{ \displaystyle 
 \int    ( tx w - x v )   dx   }\\
& & & \\
\hline
& & & \\
\quad   t^2 +  x^2  & 
{ \displaystyle 
 (  t^2 + x^2 )    { \partial \over \partial v } + 
  2t  { \partial \over \partial w } } & 
{ \displaystyle 
\sum_{ \Omega }   ( (  t^2 + x^2 )   w -  2t v )   h   }  & 
{ \displaystyle 
\int  ( (  t^2 + x^2 )   w -  2t v )  dx    }\\
& & & \\
\hline
\end{array} 
$$ 

\eject

\begin{center}
{ \bf Table~2.} Some symmetries and their generating functionals 
for the canonical bracket. The functionals 
(up to some time-dependent functions) are conservation laws 
of the semidiscrete  canonical Hamiltonian equations. 
\end{center}

$$
\begin{array}[h]{|l|l|l|l|}
\hline
{\mbox{Operator}} & 
\mbox{Operator in evolutionary form}&\mbox{Functional} & 
\mbox{Continuous limit of functional}
\\\hline
& & & \\
 { \displaystyle { \ddt}  } & 
{ \displaystyle  
{ \dh{ D } { }_0 } \left(
{ \delta  \dh {\cal{H}}  \over \delta w }  
 \right)  {\ddv}    - { \dh{ D } { }_0 } \left(
{ \delta  \dh {\cal{H}}  \over \delta v }  
 \right)  {\ddw}   }  
 &  { \displaystyle  \dh {\cal{H}}   } 
& { \displaystyle {\cal H}    } \\ 
& & & \\ \hline
& & & \\
 { \displaystyle { \ddx}  }  & 
{ \displaystyle  { \dh { \tilde{D} } { }_0 }  (v )  {\ddv} 
+ { \dh { \tilde{D} } { }_0 } (w )  {\ddw}  } 
 & { \displaystyle \sum _{\Omega } 
 w { \dh { \tilde{D} } { }_0 } ( v )  h   }
& { \displaystyle  \int  w v _x  dx    }\\ 
& & & \\ \hline
& & & \\
  { \displaystyle     } & 
{ \displaystyle  w {\ddv} -   v  {\ddw}   } 
 & 
 { \displaystyle  { 1\over 2 } \sum_{ \Omega }  
 \left( v ^2 + w ^2 \right) h } 
& { \displaystyle   { 1\over 2 } \int 
 \left( v^2 + w^2 \right) dx   }  \\ 
& & & \\ \hline
& & & \\
 { \displaystyle  }   & 
{ \displaystyle    { \dh D { }_0} ( \dh v { }_{2k} )  
{ \partial \over \partial v } +   
{ \dh D { }_0}   ( \dh w { }_{2k}  )  
{ \partial \over \partial w }   }  
 & 
 { \displaystyle \sum_{\Omega }  
w { \dh D { }_0}  ( \dh v { }_{2k}  )   h  } 
& {  \displaystyle  \int   w  v_{(2k+1)} dx    }\\
& & & \\ \hline
& & & \\
 { \displaystyle  }   & 
{ \displaystyle  \alpha ( t , x)   
{ \partial \over \partial v } + 
 \alpha _t ( t , x) 
 { \partial \over \partial w } } & 
{ \displaystyle  \sum_{ \Omega } 
( \alpha ( t , x )  w  - \alpha _t ( t , x )  v )  h } &  
{ \displaystyle  \int 
 (  \alpha ( t , x )  w  - \alpha _t ( t , x )  v ) dx }  \\
& & & \\ \hline
\end{array}
$$





\begin{thebibliography}{9999}


\bibitem{[2]}
P.~J.~Olver, { \it Applications of Lie Groups to Differential Equations}, 
Springer, New York, 1986.

\bibitem{[Kozl1]}  R.~Kozlov, 
Conservation laws of semidiscrete Hamiltonian equations, submitted to 
J. Math. Phys. 

\bibitem{[Marsd]} J.~E.~Marsden, T.~S.~Ratiu, 
{ \it Introduction to mechanics and symmetry. 
A basic exposition of classical mechanical systems},  
Texts in Applied Mathematics, 17,  Springer--Verlag, New York, 1994. 

\par
\bibitem{[3]} D.~Levi, L.~Vinet and P.~Winternitz, Lie Group Formalism 
for Difference Equations, J. Phys. A: Math. Gen. { \bf 30}, 
633--649, 1997. 


\bibitem{[Flore]} R.~Floreanini, J.~Negro, L.~M.~Nieto, L.~Vinet, 
Symmetries of the heat equation on the lattice,
 Lett. Math. Phys. {\bf 36} (1996), no. 4,
351--355, 1996.


\bibitem{[Vinet]} R.~Floreanini, L.~Vinet, 
Lie symmetries of finite--difference equations, 
J. Math. Phys. { \bf 36}, no. 12, 7024--7042, 1995.


\bibitem{[5c]}
V.~A.~Dorodnitsyn, Transformation groups in a space of difference variables,
in  VINITI Acad. Sci. USSR, Itogi Nauki i Techniki,  
{\bf {34}}, 149--190, 1989,
(in Russian), see English translation in J. Sov. Math.  
{\bf {55}}, 1490, 1991.

\bibitem{[5b]}
V.~A.~Dorodnitsyn, Newton's group and commutative properties of
Lie--B\"{a}cklund operators in finite difference space, Preprint of Keldysh
Institute of Applied Mathematics,  no. 175, Moscow,  1988, (in Russian). 




\bibitem{[Sanz]} J.~M.~Sanz-Serna,  M.~P.~Calvo, 
{ \it Numerical Hamiltonian problems}, 
Applied Mathematics and Mathematical Computation, 7, 
Chapman $\&$ Hall, London, 1994. 


\par
\bibitem{[Brid1]}  Th.~J.~Bridges, 
Multi--symplectic structures and wave propagation, 
Math. Proc. Cambridge Philos. Soc. {\bf 121}, no. 1, 147--190, 1997.

\par
\bibitem{[Reich1]} S.~Reich, 
Multi--symplectic Runge-Kutta collocation methods 
for Hamiltonian wave equations, J. Comput. Phys. {\bf 157},
473--499, 2000.

\par
\bibitem{[Reich2]} Th.~J.~Bridges and S.~Reich, 
Multi--symplectic integrators: numerical schemes 
for Hamiltonian PDEs that conserve symplecticity, Technical report. 

\par
\bibitem{[Reich3]} Th.~J.~Bridges and S.~Reich, 
Multi--symplectic spectral discretizations for 
the Zakharov-Kuznetsov and shallow water equations, Technical report. 






\bibitem{[72]}
W.~F.~Ames, R.~L.~Anderson, V.~A.~Dorodnitsyn, E.~V.~Ferapontov, 
R.~K.~Gazizov, N.~H.~Ibragimov and S.~R.~Svirshchevskii, 
{\it CRC Hand-book
of Lie Group Analysis of Differential Equations, 
ed. by N.~Ibragimov, Volume I: Symmetries,
Exact Solutions and Conservation Laws}, CRC Press, 1994.



\par
\bibitem{[71]}
A.~V.~Aksenov, V.~A.~Baikov, V.~A.~Chugunov, R.~K.~Gazizov, A.~G.~Meshkov, 
{\it CRC Hand-book
of Lie Group Analysis of Differential Equations, 
ed. by N.~Ibragimov, Volume II: Applications in engineering and 
physical sciences}, CRC Press, 1995.



\par
\bibitem{[Zah]} V.~E.~Zaharov, 
{ \it Handbook of Plasma Physics, Vol.~2, edited 
by M.~N.~Rosenbluth and R.~Z.~Sagdeev}, New York, 1984.  


\bibitem{[10]} N.~H.~Ibragimov, 
{\it Transformation groups applied to mathematical physics},  
Dovdrecht, D.~Reidel, 1995. 




\end{thebibliography}
\end{document}